\documentclass[aps,twocolumn,superscriptaddress,nofootinbib]{revtex4}
\usepackage{amsmath}
\usepackage{graphicx}
\usepackage{caption}
\usepackage{color}
\usepackage{slashed}

\begin{document}

\title{Charge-mass ratio bound and optimization in the Parikh-Wilczek tunneling model of Hawking radiation}

\author{Kyung Kiu Kim}\thanks{%
E-mail: kimkyungkiu@gmail.com}
\affiliation{Department of physics, Kyung Hee University, Seoul, Korea}
\affiliation{Center for Quantum Space Time, Sogang University, Seoul, Korea}

\author{Wen-Yu Wen}\thanks{%
E-mail: steve.wen@gmail.com}

\affiliation{Department of Physics and Center for High Energy Physics, Chung Yuan Christian University, Chung Li City, Taiwan}
\affiliation{Leung Center for Cosmology and Particle Astrophysics\\
National Taiwan University, Taipei 106, Taiwan}

\begin{abstract}
In this letter, we study the mutual information hidden in the Parikh-Wilczek tunneling model of Hawking radiation for Reissner-Norstr\"{o}m black holes.  We argue that the condition of nonnegativity of mutual information suggests bound(s) for charge-mass ratio of emitted particles.  We further view the radiation as an optimization process and discuss its effect on time evolution of a charged black hole.
\end{abstract}

%\pacs{11.25.Tq, 74.20.-z}
\maketitle

%\date{\today}

%\keywords{}

\section*{Parikh-Wilczek tunneling model of Hawking radiation}
\begin{figure}[tbp]

\includegraphics[width=0.45\textwidth]{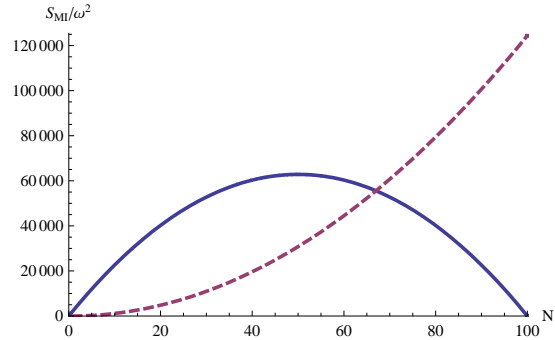} 
\caption{\label{fig:vonNeumann}Consider a black hole $M$ evaporates in $N$ steps with a particle of equal mass $\omega$ each time.  The mutual information between the black hole and emitted particles (solid curve) first increases then decreases, while total mutual information between pairs of emitted particles (dashed curve) increases quadratically .  Here we use $M/\omega=100$ to simulate the result.}

\end{figure}

The original treatment of Hawking radiation by Hawking is to consider perturbation in a fixed background of Schwarzschild black hole.  The thermal spectrum brought up controversial debates over the Information Loss Paradox.  Parikh and Wilczek considered radiation as an outgoing tunneling particle where the conservation of energy is enforced\cite{Parikh:1999mf}.  For the Schwarzschild black holes of mass $M$ and radiation $\omega$, the tunneling probability reads $\Gamma(M,\omega)\sim \exp{[-8\pi \omega(M-\frac{\omega}{2})]}$\footnote{In this letter, we will adopt the natural units such that $G=c=4\pi\epsilon_0=1$.}.  The radiation is obviously not thermal because two consecutive emmisions are not independent, that is $\Gamma(M,\omega_1+\omega_2)\neq \Gamma(M,\omega_1)\cdot \Gamma(M,\omega_2)$.  In other words, the latter emission depends on the previous one such that the conditional probability $\Gamma(M,\omega_2 | \omega_1)\neq \Gamma(M,\omega_2)$, where $\Gamma(M,\omega_2 | \omega_1)\equiv\Gamma(M-\omega_1,\omega_2)$.  The logarithmic difference between two quantities defines a mutual information or correlation between two consecutive emssion\cite{Zhang:2009jn}:
\begin{equation}
S_{MI}(M,\omega_2:\omega_1)\equiv S(M,\omega_2|\omega_1)-S(M,\omega_2),
\end{equation}
where we define the entropy function $S(M,\omega_i)=\ln \Gamma(M,\omega_i)$ and $S(M,\omega_i|\omega_j)=\ln \Gamma(M,\omega_i|\omega_j)$.  It is obvious to see from definition that the mutual information vanishes if two emissions are independent.  A simple evaluation yields $S_{MI}(M,\omega_2:\omega_1)=8\pi \omega_1 \omega_2$ for the tunneling model\cite{Zhang:2009jn}.  Using this definition of mutual information, we show in the figure \ref{fig:vonNeumann} that by considering pairwise entanglement between Schwarzschild black hole and emitted particles.  It is entertaining to compare with an earlier unitary model of black hole evaporation proposed by Page\cite{Page:1993wv} and recently by Iizuka and Kabat\cite{Iizuka:2013ria} .

In fact, the tunneling probability takes a universal form for many kinds of black hole with finite size of event horizon: $\Gamma\sim e^{\Delta S_{BH}}$, where $\Delta S_{BH}$ is the change of Bekenstein-Hwaking entropy after radiation.  In particular, the tunneling model of neutral particles in the Reissner-Norstr\"{o}m black hole was discussed in Parikh and Wilczek's original work\cite{Parikh:1999mf} and it was generalized to charged particles later in \cite{Zhang:2005xt}.  In the case of the Reissner-Norstr\"{o}m black holes with mass $M$ and charge $Q$, we denote the entropy function as $S(M,Q;\omega,q)=\pi\{[(M-\omega)+\sqrt{(M-\omega)^2-(Q-q)^2}]^2-\big(M+\sqrt{M^2-Q^2} \big)^2\}$ for each emission of mass $\omega$ and charge $q$.  In order to avoid naked singularity, condition $Q \le M$ has to be satisfied during the Hawking radiation.  This suggests the existence of certain bound(s) for charge-mass ratio in each emission.  In the original treatment of Hawking radiation, it is not obvious how to estimate this ratio bound while the conservation of energy is not enforced.  On the other hand, while the conservation law is enforced in the tunneling model, the conservation of information is also guaranteed\cite{Zhang:2009jn}.  Whether the Information Loss Paradox is solved or not in the tunneling model by admitting nontrivial entanglement during the radiation process is still under dedate\cite{Mathur:2011uj,Cai:2012um}.  Regardless the consequence of debate, our main results in this letter are to estimate the charge-mass ratio bound from the mutual information shared by two consecutive emissions, and propose an optimization scheme in the radiation process.

\section*{Nonnegativity condition of mutual information}

Since the monotonically decreasing of black hole entropy during Hawking radiation, it is a natural assumption that mutual information stored in each consecutive emission pair is nonnegative.  This fact is known as the Jensen's inequality from the mathematical point of view\footnote{A similar fact in the probability theory states that the probability for event A to happen under condition B is no less than the probability for event A to happen with no condition, i.e. $P(A|B) = P(A\cap B)/P(B) \ge P(A)P(B)/P(B) = P(A)$.}.  The nonnegativity condition for the mutual information in radiation of Schwarzschild black holes simply requires that $\omega_i\geq 0$ in each emission, which is just the nonnegativity of mass or energy\footnote{In the case of Schwarzschild black hole, the nonnegativity condition of mutual information is simply due to the fact that the entropy function $S(M,\omega)$ defined earlier is a monotanously decreasing function with respect to $M$, that is $\partial S(M,\omega) / \partial M = -8\pi \omega < 0$.  Nevertheless, the nonnegativity condition is not automatically satisfied in the case of RN black hole because the $S(M,Q;\omega,q)$ is not a simple function when both $M$ and $Q$ are varying.}.  Investigating the mutual information appeared in the Reissner-Norstr\"{o}m black hole suggests a bound for charge-mass ratio.  The mutual information between two consecutive emissions with mass $\omega_1, \omega_2$ and charge $q_1,q_2$ can be computed as $S_{MI}(M,Q;\omega_2,q_2:\omega_1,q_1)\equiv S(M,Q;\omega_2,q_2|\omega_1,q_1)-S(M,Q;\omega_2,q_2)$.  While this quantity is nontrivially dependent on $M$ and $Q$, it is insightful to obtain some simple results at following limits:
\begin{itemize}
\item  $M \gg \omega_1,\omega_2$, $Q \gg q_1,q_2$ and $M \gg |Q|$\\
In the limit of large black mass and charge, we have a simple form: $S_{MI}(M,Q;\omega_2,q_2:\omega_1,q_1)=4\pi (2\omega_1 \omega_2-q_1q_2)$.  If one assumes that two emissions are identical for simplicity, the nonnegativity condition enforces an upper bound for the charge-mass ratio $|q_i| /\omega_i \le \sqrt{2}$.

\item  $M \approx |Q| \gg \omega,q $\\
In this near extremal and large mass(charge) limit, one finds the nonnegativity condition gives a small window for possible charge-mass ratio, that is  $1-\epsilon_- \leq |q|/\omega \leq 1+\epsilon_+$ for $\epsilon_\pm \sim {\cal O}(10M^{-1})$.

\item	 $M=2\omega,Q=2q$ \\
In this last stage of black hole evaporation, the nonegativity condition that $S=4\pi \omega^2 - 2\pi q^2 + 4 \pi \omega \sqrt{\omega^2-q^2} \geq 0$ implies an upper bound $|q|/\omega \leq 1$.

\end{itemize}

\begin{figure}[t]
\includegraphics[width=0.5\textwidth]{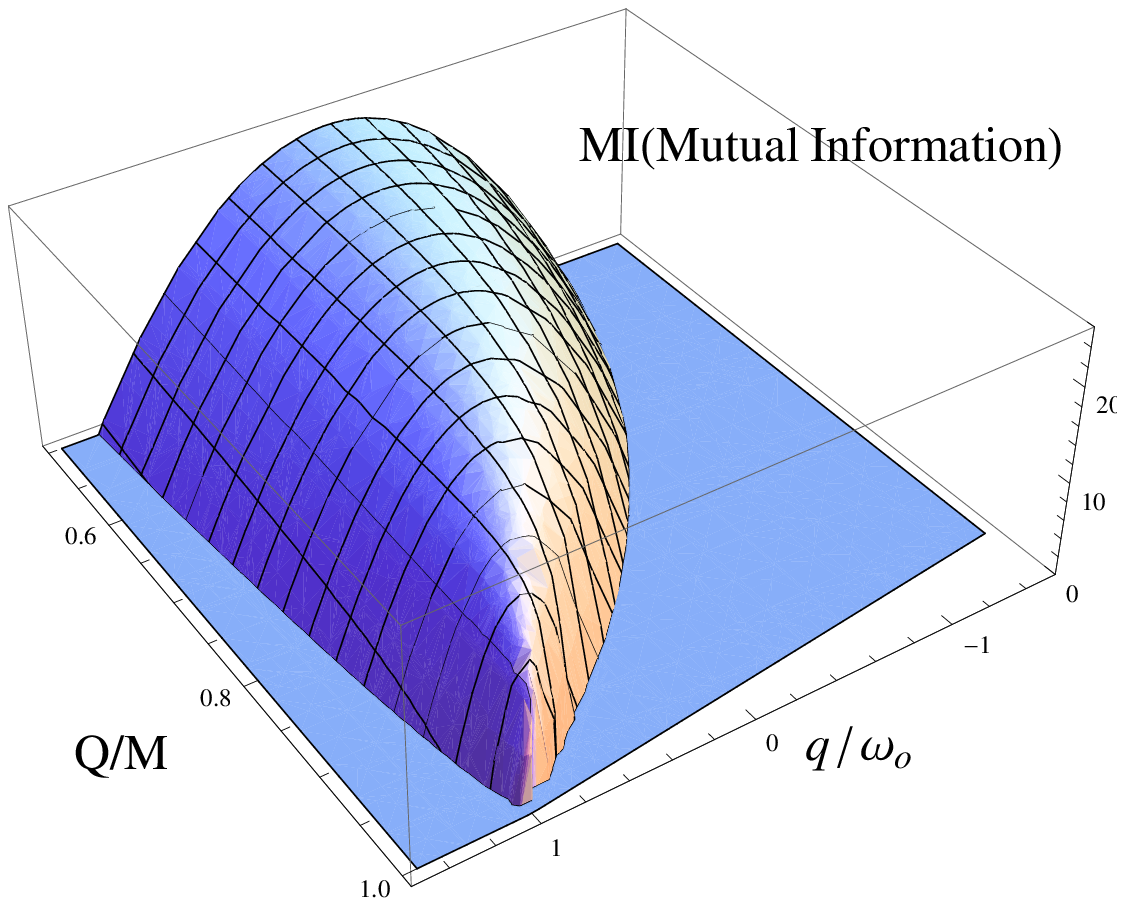} 
\caption{\label{fig:ratio_bound}The charge-mass ratio bound for various ratios $|Q|/M$.  In order to easily show the location of maximum, we also plot the exclusive region of ngative ratio.  The lower bound appears when $Q/M > 0.86$.}

\includegraphics[width=0.4\textwidth]{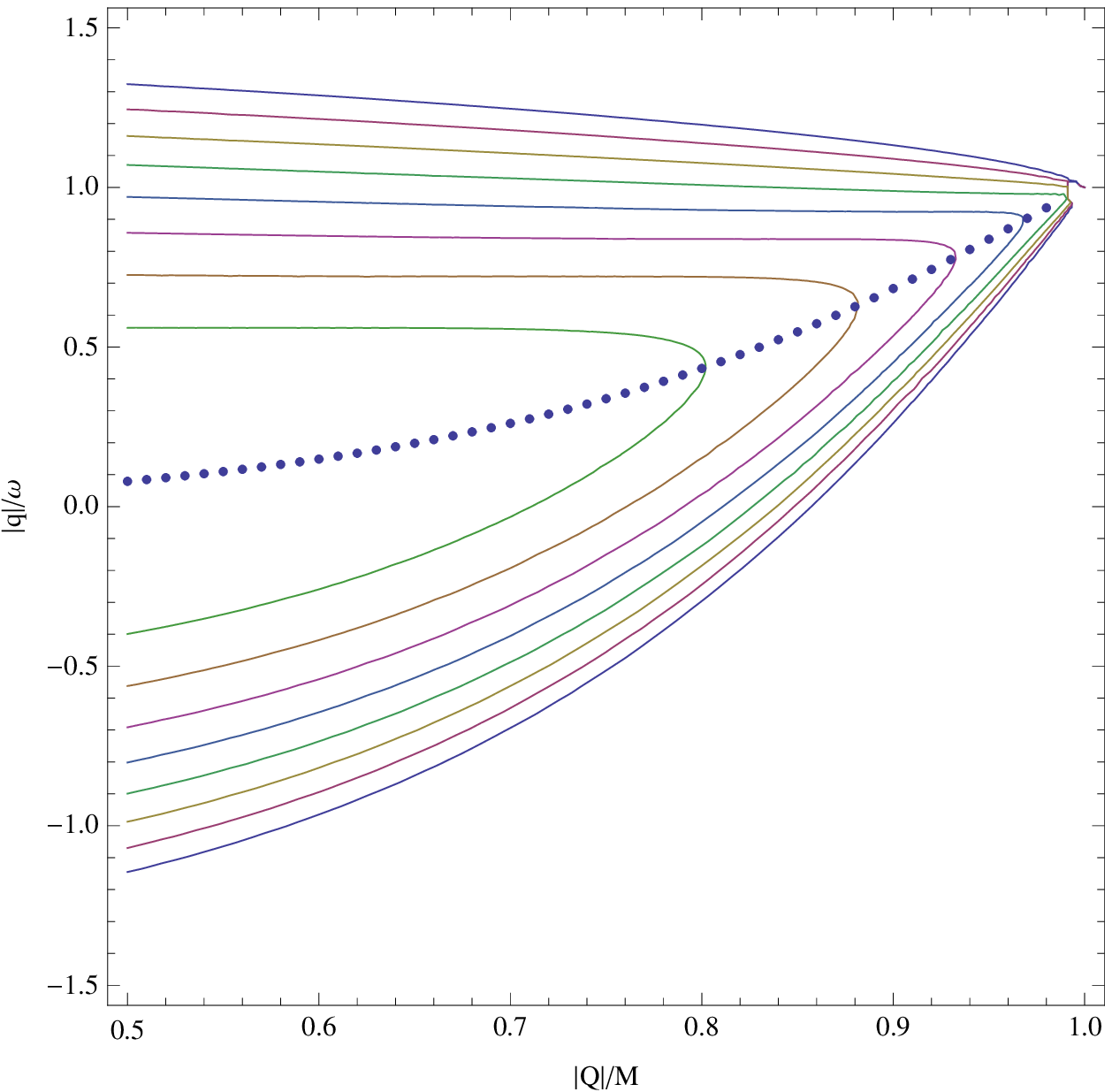} 
\caption{\label{fig:loci}The contour plot of figure \ref{fig:ratio_bound}.  The ratio $|q|/\omega$ of maximum mutual information (blue dots) changes from $0$ to $1$ while $|Q|/M$ approaches extremality.  The color contours show the equi-height lines of mutual information with values between $0$ and $21$.}
\end{figure}

Apart from those above-mentioned limits, one can numerically show the existence of bound(s) is universal for various values of black hole mass and charge.  As shown in the figure \ref{fig:ratio_bound}, the upper bound is between $\sqrt{2}$ and $1$ and there also appears nonzero lower bound near extremal limit while $Q/M > 0.86$.

\subsection*{Maxmimum mutual information optimization}

We have learnt that the emitted quanta have bounds for the charge-mass ratio, but it does not tell exactly how much information is carried away during each emission.  To estimate the amount of information for each emission, we propose the following alternative mechanism for evaporation of a charged black hole:

\begin{itemize}

\item
Given a specific mass and charge of a charged black hole, emissions of nonnegative mutual information are all admitted with some probability.

\item
The emission carrying more mutual information has more chance.  That is, the emission with maximum mutual information (MMI) dominates the process\footnote{We remark that idea of MMI has played an important role in the signals transmission in a neural system of multiple inputs and ourputs\cite{linsker}.}.

\end{itemize}

It is possible to realize above-mentioned mechanism in the language of path integral, if an action relevant to the mutual information could be assigned to different paths.  From information point of view, this is an optimization process that a charged black hole radiates most efficiently by giving away as much information as possible.  The optimized $|q|/\omega$ ratio of emission with MMI can be estimated by assuming two consecutive emissions are small.  In the figure \ref{fig:ratio_bound} and \ref{fig:loci}, we plot the function $S_{MI}$ and loci of MMI for various $|Q|/M$ ratio and emitted $|q|/\omega$ ratio.  Several remarks are in order:

\begin{itemize}

\item
The mutual information, carried away by pair of emissions with mass $\omega$, has maximum value $8\pi \omega^2$ for the neutral Schwarzschild black hole.

\item
For the charged black hole, the mutual information decreases with the $|Q|/M$ ratio.  That is, the closer to the extremal limit, the less information {\sl leaks} via radiation.  The process is expected to stop at the extremal limit due to vanishing mutual information.  However, it might take infinite time (steps) to reach extremal limit from non-extremality as discussed in \cite{Fabbri:2000xh}.

\item
The charge-mass ratio of MMI, denoting $\gamma_M$, is given by
\begin{equation}
\gamma_M = \frac{Q^3}{M^3+(M^2-Q^2)^{3/2}},
\end{equation}
which is very small while the charged black hole is far away from extremal limit, and approaches unity while it is near extremality.

\end{itemize}

\begin{figure}[tbp]

\includegraphics[width=0.5\textwidth]{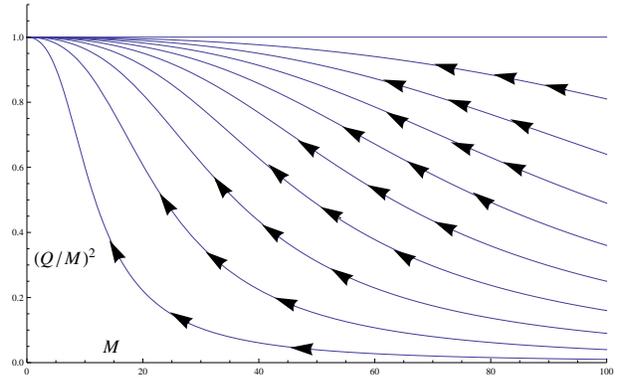} 
\caption{\label{fig:evaporate}Evaporation under the MMI optimization.  The process starts with some initial ratio $0<|Q|/M<1$ and leads to the final state $(M,|Q|/M)=(0,1)$, where a black hole vanishes at zero temperature.}

\end{figure}

\begin{figure}

\includegraphics[width=0.5\textwidth]{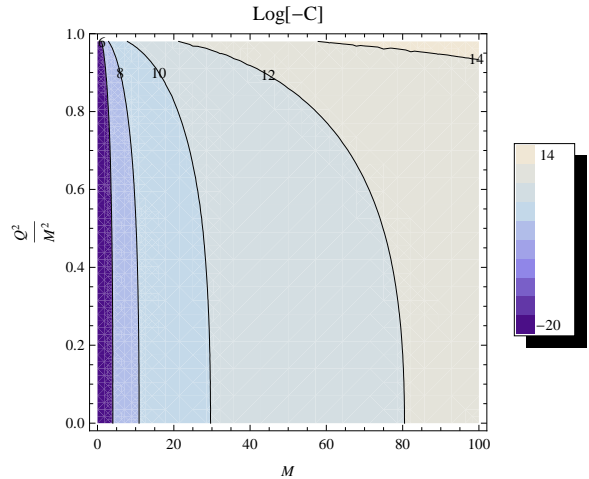}
\caption{\label{fig:specific_heat}The contour plot of $Log(-C)$.  The absolute value $|C|$ grows with black hole mass and get larger far away from extremality.}
\end{figure}

The evaporation of charged black holes were studied in detail by Hiscock and Weems\cite{Hiscock:1990ex}, where the radiation spectrum is thermal and Schwinger formula for pair-production were used for charge dissipation.  The evaporation process respecting the MMI optimization can be simulated for charged black holes with various initial masses $M$ and charges $Q$.  As shown in the figure \ref{fig:evaporate}, it seems to roughly agree with the mass dissipation zone in \cite{Hiscock:1990ex} at the early stage, however at later stage, it leads to the extremal limit before mass is completely exhausted, while it leads farther away from extremality in Hiscock and Weems.  There appears no charge dissipation zone in our model.   These differences can be also understood by explicitly evaluating the specific heat in our model:
\begin{equation}
C \equiv \frac{dM}{dT} = \frac{2\pi r_+^4 (Q^2+M(r_+-3M))}{(M^2-Q^2)(2M r_+ -Q^2)}.
\end{equation}
We plot the specific heat for various $M$ and ratio $Q/M$ in the figure \ref{fig:specific_heat}.  In constrast to that in \cite{Hiscock:1990ex}, the specific heat in our model is always negative, but approaching zero before it completely evaporates at extremality.  We remark that to derive this specific heat, one noly needs to impose the optimization of MMI, without knowing details about charge and mass dissipation.

\section*{Discussion}
The charge-mass ratio bound is found in this letter to be of order unity, that is about $\sqrt{\frac{G}{4\pi \epsilon_0}}= 8.17\times 10^{-11} C/kg$.
It is worth mentioning that similar charge-mass ratio bound near the charged black hole was investigated by including the self-interaction of charged particles due to the infinite redshift near black hole horizon\cite{Hod:2010zk}, and recently by one of the authors using the tunneling method\cite{Wen:2013zcx}.  Based on the same tunneling model, the ratio bound imposed by the nonnegativity of mutual information is tighter than that by the self-force\cite{Wen:2013zcx}, nevertheless both bounds approaches unity when the black hole is near the extremality.  On the other hand, it would be interesting to check whether the evaporation scenario proposed in \cite{Hiscock:1990ex} respects the nonnegativity constraint.  Since we have learnt that in the tunneling model, the nonnegative mutual information is in fact a direct consequence of monotoneous decreasing of black hole entropy during evaporation of both Schawrzschild and Reissner-Nordstr\"{o}m black holes, it can serve as a guiding principle to rule out any model with unphysical process.  Furthermore, as we show in this paper, it can even {\sl dictate} the dynamics if optimization is called.

\begin{acknowledgments}
We are grateful to useful discussion with Dr. Dong-han Yeom.  WYW is supported in part by the Taiwan's National Science Council under grant NO. 102-2112-M-033-003-MY4 and the National Center for Theoretical Science. KKK is supported by the National Research
Foundation of Korea(NRF) grant funded by the Korea government(MEST) with
grant No. 2010-0023121, 2011-0023230, 2012046278, and also through the Center
for Quantum Spacetime(CQUeST) of Sogang University with grant number 2005-
0049409.
\end{acknowledgments}

%%%%%%%%%%%%%%%%%%%%%%%%%%%%%%%%%%%%%%%%%%%%%%%%%%%%%%%%%%%%%%%

\end{document}